\documentclass[11pt]{article}

\usepackage{amssymb}
\usepackage{amsmath}
\usepackage{graphicx}
\def\beq{\begin{eqnarray}}    
\def\eeq{\end{eqnarray}}      

\newcommand{\OM}{\Omega_M}
\newcommand{\OL}{\Omega_{\Lambda}}

\newcommand{\OK}{\Omega_K}

\newcommand{\rco}{\rho_{c,0}}
\newcommand{\rs}{\rho_s}

\newcommand{\rM}{\rho_M}

\newcommand{\rD}{\rho_D}
\newcommand{\rL}{\rho_{\CC}}
\newcommand{\pD}{p_D}
\newcommand{\wD}{\omega_D}
\newcommand{\CC}{\Lambda}

\newcommand{\we}{\omega_{e}}
\newcommand{\tOM}{\tilde{\Omega}_M}

\newcommand{\tOD}{\tilde{\Omega}_{D}}
\newcommand{\xL}{\xi_{\CC}}
\newcommand{\fM}{f_{M}}
\newcommand{\fL}{f_{\Lambda}}
\begin{document}
\hyphenation{cos-mo-lo-gi-cal sig-ni-fi-cant}



\begin{center}
{\large \textsc{Dynamical dark energy or variable cosmological
parameters?}} \vskip 2mm

 \vskip 8mm

\textbf{Joan Sol\`{a}}$^{a,b}$,\ \textbf{Hrvoje
 \v{S}tefan\v{c}i\'{c}}$^{a}$\footnote{On leave of absence from
the Theoretical Physics Division, Rudjer Bo\v{s}kovi\'{c}
Institute, Zagreb, Croatia.} \vskip0.5cm $^{a}$ {HEP Group, Dep.
ECM, Universitat de Barcelona\\  Av. Diagonal 647, 08028
Barcelona, Catalonia, Spain}

$^{b}$  C.E.R. for Astrophysics, Particle Physics and
Cosmology\,\footnote{Associated with Institut de Ci\`encies de
l'Espai-CSIC.}

E-mails: sola@ifae.es, stefancic@ecm.ub.es

\vskip2mm

\end{center}
\vskip 15mm

\begin{quotation}
\noindent {\large\it \underline{Abstract}}.$\,\,$ One of the main
aims in the next generation of precision cosmology experiments
will be an accurate determination of the equation of state (EOS)
for the dark energy (DE). If the latter is dynamical, the
resulting barotropic index $\omega$ should exhibit a non-trivial
evolution with the redshift. Usually this is interpreted as a
sign that the mechanism responsible for the DE is related to some
dynamical scalar field, and in some cases this field may behave
non-canonically (phantom field). Present observations seem to
favor an evolving DE with a potential phantom phase near our
time. In the literature there is a plethora of dynamical models
trying to describe this behavior. Here we show that the simplest
option, namely a model with a variable cosmological term,
$\Lambda=\Lambda(t)$, leads in general to a non-trivial effective
EOS, with index $\we$, which may naturally account for these data
features. We prove that in this case there is \textit{always} a
``crossing'' of the $\we=-1$ barrier near our time. We also show
how this effect is modulated (or even completely controlled) by a
variable Newton's gravitational coupling $G=G(t)$.
\end{quotation}
\vskip 8mm

\newpage

\vskip 6mm

 \noindent {\bf 1. Introduction}\quad

 \vskip 0.4cm

Cosmology is becoming a science of precision and accuracy. In the
last few years a flurry of experimental activity has been devoted
to the measurement of the cosmological parameters. This effort has
transformed observational cosmology into a fairly respectable
branch of experimental physics\,\cite{Supernovae,WMAP03,LSS}. As a
result, evidence is piling up in favor of the existence of the
dark energy (DE) component, $\rD$, pervading the Universe (or at
least the known patch of it). Hopefully, this situation is bound
to improve further with the advent (in the near future) of the
promising SNAP and PLANCK projects \,\cite{SNAP}, which will grant
an experimental determination of the cosmological parameters to an
unprecedented few percent level of accuracy. Quite in contrast,
the nature of the DE remains at the moment a profound mystery.
Historically, the DE was first identified with the cosmological
constant (CC), $\Lambda$, and the vacuum energy contributions it
receives from quantum field theory (QFT)\,\cite{zeldo}. This idea
led to the famous cosmological constant
problem\,\cite{weinRMP,CCRev}. In more recent times the notion of
DE has been detached from that of $\Lambda$ and has been extended
to a variety of models leading to an accelerated expansion of the
universe in which the DE itself is a time-evolving entity. These
models include dynamical scalar fields (quintessence and the
like)\cite{Wetterich,Peebles}, phantom fields\,\cite{phantom},
braneworld models\,\cite{braneworld}, Chaplygin
gas\,\cite{Chaplygin}, and many other recent ideas like
holographic dark energy, cosmic strings, domain walls etc (see
e.g.  \,\cite{CCRev} and references therein). Obviously the very
notion of CC in such broader context becomes diminished, the CC
could just be inexistent or simply relegated to the status of one
among many other possible candidates.

A general DE model is described as being a sort of fluid
characterized by a conserved DE energy density, $\rho_D$, and
pressure $p_{D}$, related by an equation of state (EOS)
$p_{D}=\wD\,\rho_D$. A particular and popular realization of the
DE is the aforementioned \textit{quintessence}
idea\,\cite{Wetterich,Peebles}, where one has some scalar field
$\chi$ which generates a non-vanishing $\rho_D$ from the sum of
its potential and kinetic energy term at the present time:
$\rho_D^0=\{(1/2)\xi\,\dot\chi^2+V(\chi)\}_{t=t_0}$. The sign of
the coefficient $\xi$ determines whether the field can describe
quintessence ($\xi>0$) or phantom DE ($\xi<0$). If the kinetic
energy for $\chi$ is small enough, it is clear that $\rho_D^0$
looks as an effective cosmological constant. The scalar field
$\chi$ is in principle unrelated to the Higgs boson or any other
field of the Standard Model (SM) of particle physics, including
all of its known extensions (e.g. the supersymmetric
generalizations of the SM); in other words, the $\chi$ field is an
entirely \textit{ad hoc} construct just introduced to mimic the
cosmological term. However, the field $\chi$ is usually thought of
as a high energy field (unrelated to SM physics), i.e. $\chi\simeq
M_X$ where $M_X$ is some high energy scale typically around the
Planck mass $M_P\sim 10^{19}\,GeV$. If one assumes the simplest
form for its potential, namely
$V(\chi)=(1/2)\,m_{\chi}^2\,\chi^2$, its mass turns out to be
$m_{\chi}\sim\,H_0\sim 10^{-33}\,eV$ in order to approximately
describe the present value of the energy density associated to
$\CC$, {namely $\rL^0\equiv\CC_0/8\pi\,G\sim 10^{-47}\,GeV^4$}.
Obviously it is very difficult to understand the small mass
$m_{\chi}$ in particle physics, and this is of course a serious
problem underlying the quintessence models. On the other hand the
possible observed transition from a quintessence into phantom-like
behavior\,\cite{Alam,Jassal} cannot be explained with only one
scalar field (because the sign of the coefficient $\xi$ is fixed),
and therefore one has to resort to at least two scalar fields, one
with $\xi>0$ and the other with another coefficient
$\xi'<0$\,\,\footnote{For various recent theoretical approaches to
the $\wD = -1$ boundary crossing see e.g. \cite{cross}.}. In spite
of the many difficulties with the scalar field models they have
the virtue that they may help to understand why the DE might be
evolving with time as suggested by the analyses of the
cosmological data\,\cite{Alam,Jassal}, and in this sense one would
like not to lose this useful aspect of the quintessence proposal.

In this work, in contrast to the previous approaches, we wish to
stick to the original idea that the primary cause of the DE is a
``true'' CC term in Einstein's equations, but we assume it is a
variable one: $\CC=\CC(t)$. Actually, a rich variety of variable
$\CC$ models have been reported in the literature -- some of them
admittedly on purely phenomenological
grounds\,\cite{CCvariable1,CCvariable2}. There are however models
based on a more fundamental premise, even if ultimately
phenomenological at the moment. For instance, there are models
attempting to adjust the small value of the cosmological term with
the help of a dynamical scalar field\,\cite{Dolgov}, as for
example the \textit{cosmon} model \,\cite{Wetterich,PSW}. More
recently a variable $\CC$ has been proposed as a running quantity
from the point of view of the renormalization group
(RG)\,\cite{JHEPCC1}-\cite{Reuter}. Particularly interesting for
the present work is the possibility that an effective equation of
state can be associated to a variable cosmological model. This has
been explored in detail in Ref.\cite{SS1} for a model of running
cosmological term\,\cite{RGTypeIa}. In the present work we
generalize this idea to arbitrary variable cosmological models
(without specifying the underlying dynamics), and show that these
models can effectively lead to a non-trivial EOS (therefore
mimicking a dynamical DE model). We also show how to include in
the effective EOS the possible effects from a variable Newton's
gravitational coupling. We provide a general algorithm to
construct this effective EOS and prove some general results which
describe the relationship between the variable $\CC$ models and
the dynamical DE models. Finally we argue that this relationship
may help to shed some light to understand the apparently observed
transition from quintessence-like to phantom-like behavior as
indicated by some analyses of the most recent cosmological
data\,\cite{Alam,Jassal}.

\vspace{0.7cm}

\noindent {\bf 2. Two cosmological pictures: Variable {$\CC$ and
$G$} versus dynamical dark energy}\quad \vskip 0.4cm

We start from Einstein's equations in the presence of the
cosmological constant term,
\begin{equation}
R_{\mu \nu }-\frac{1}{2}g_{\mu \nu }R=8\pi
G\,T_{\mu\nu}+\,g_{\mu\nu}\CC\equiv 8\pi
G\,(T_{\mu\nu}+g_{\mu\nu}\,\rL)\,, \label{EE}
\end{equation}
where $T_{\mu\nu}$ is the ordinary energy-momentum tensor
associated to isotropic matter and radiation, and $\rL$
represents the energy density associated to the CC.  Let us next
contemplate the possibility that $G=G(t)$ and $\CC=\CC(t)$ can be
both functions of the cosmic time within the context of the FRW
(Friedmann-Robertson-Walker) cosmology. The possibility of a
variable gravitational coupling and a variable cosmological term
in cosmology (variable CC term, for short) has been considered by
many authors, and in particular it has been discussed within the
renormalization group approach to
cosmology\,\cite{JHEPCC1}-\cite{Reuter}. It should be clear that
the very precise measurements of $G$ existing in the literature
refer only to distances within the solar system and astrophysical
systems. In cosmology these scales are immersed into much larger
scales (galaxies and clusters of galaxies) which are treated as
point-like (and referred to as ``fundamental observers'',
co-moving with the cosmic fluid\,\cite{Rindler}). Therefore, the
variations of $G$ at the cosmological level could only be seen at
much larger distances where we have never had the possibility to
make direct experiments. To put it in another way, the potential
variation of $G=G(t)$ and $\CC=\CC(t)$ is usually tested in terms
of a measurable redshift dependence of these functions, $G=G(z)$
and $\CC=\CC(z)$, say a variation for a redshift interval of at
least $\Delta z>0.01$, therefore implying scales of several
hundred Mpc.

The only possible EOS for the CC term, whether strictly constant
or variable, is $p_{\CC}=-\rL$. However, we may still parametrize
this variable $\CC$ model as if it would be a dynamical field DE
model with pressure and density $(\pD,\rD)$. We will call this
the ``effective DE picture'' of the fundamental $\CC$ model. The
matching of the DE picture with the fundamental CC picture
generates a non-trivial ``effective EOS''\,\cite{Eqos} for the
latter, $\pD=\we\,\rD$, where the effective barotropic index
$\we=\we(z)$ is a function of the cosmological redshift to be
determined.  By exploring this alternative we can test the
variable $\CC$ models as a canonical source of effective
dynamical DE models. In the following we will restrict our
considerations to the flat space case ($\OK=0$). This is not only
for the sake of simplicity, but also because it is the most
favored scenario at present. Friedmann's equation with
non-vanishing $\rL$ reads
\begin{equation}\label{CCpicture}
H_{\CC}^2=\frac{8\pi G}{3}(\rho+\rL)\,,
\end{equation}
where for convenience we have appended a subscript in the Hubble
parameter. On the other hand the general Bianchi identity of the
Einstein tensor in (\ref{EE}) leads to
\begin{equation}\label{GBI}
\bigtriangledown^{\mu}\,\left[G\,(T_{\mu\nu}+g_{\mu\nu}\,\rL)\right]=0\,.
\end{equation}
Using the FRW metric explicitly, the last equation results into
the following ``mixed'' local conservation law:
\begin{equation}\label{BianchiGeneral}
\frac{d}{dt}\,\left[G(\rho+\rL)\right]+3\,G\,H_{\CC}\,(\rho+p)=0\,.
\end{equation}
If {$\dot\rL\equiv d\rL/dt\neq 0$}, $\rho$ is not generally
conserved as there may be transit of energy from matter-radiation
into the variable $\rL$ or vice versa (including a possible
contribution from a variable $G$, if $\dot{G}\neq 0$). Thus this
law indeed mixes the matter-radiation energy density with the
vacuum energy ($\rL$). {To be more precise, the following
scenarios are possible: i) $G=$const. \textit{and} $\rL=$const.
This is the standard case of $\CC$CDM cosmology implying the
canonical conservation law of matter-radiation:
$\dot{\rho}+3\,H\,(\rho+p)=0$, ii) $G=$const \textit{and}
$\dot{\rL}\neq 0$, in which case Eq.(\ref{BianchiGeneral}) boils
down to $\dot{\rL}+\dot{\rho}+3\,H\,(\rho+p)=0$; iii) $\dot{G}\neq
0$ \textit{and} $\rL=$const, implying
$\dot{G}(\rho+\rL)+G[\dot{\rho}+3H(\rho+p)]=0$; and finally iv)
$\dot{G}\neq 0$ \textit{and} $\dot{\rL}\neq 0$, which in the case
of self-conservation of matter-radiation it leads to
$(\rho+\rL)\dot{G}+G\dot{\rL}=0$. Notice that in cases ii) and
iii) the matter-radiation cannot be canonically conserved (if one
of the two parameters $\rL$ or $G$ indeed is to be variable),
whereas in case iv) it is assumed to be conserved. Explicit
cosmological models with variable parameters as in  cases ii) and
iv) have been constructed within the context of RG cosmology in
Ref.\cite{RGTypeIa} and \cite{SSS} respectively. }

{With only Eq.(\ref{BianchiGeneral}) and the FRW equations we
cannot solve the cosmological model with variable parameters. One
needs a fundamental model that informs us on the functional
dependence of $\rL$ and $G$ on the remaining cosmological
functions, like $\rho, p$ and $H$. However, in most of this work
we will not commit ourselves to any specific model for the
underlying fundamental dynamics, because the kind of results we
are aiming at are to be valid for a large class of models with
dynamical cosmological parameters. Our guiding paradigm will be
the general expectation that these parameters are variable because
of the effective running that they should display according to the
renormalization group. In this sense the possible variation of
``fundamental constants'' such as $\CC$ and $G$ could be an
effective description of some deeper dynamics associated to QFT in
curved space-time, or quantum gravity or even string theory, all
of which share the powerful RG approach to quantum effects. For
instance, in the specific framework of QFT in curved space-time
the RG equations for $\rL$ and $G$ should entail definite laws
$\rL=\rL(t)\,, G=G(t)$}\,\cite{Book}. {Given a fundamental model
based on QFT}, the parameter $\rL$ will primarily depend on some
cosmological functions (matter density $\rho$, Hubble expansion
rate $H$, etc) which evolve with time or redshift. Similarly for
the Newton coupling $G$. Therefore, in general we will have two
functions of the redshift
\begin{equation}\label{varibleCCG}
\rL(z)=\rL(\rho(z),H(z),...)\,,\ \ \ \ \
G(z)=G(\rho(z),H(z),...)\,.
\end{equation}
{We understand that other fundamental parameters could also be
variable. For example, the fine structure constant has long been
speculated as being potentially variable with the cosmic time
(and therefore with the redshift) including some recent
experimental evidences -- see e.g. \,\cite{Webb}. This might also
have an interpretation in terms of the RG at the level of the
cosmological evolution. And similarly with other ``constants''
such as the ratio $m_e/m_p$ between the electron to the proton
mass etc which could also have evolved throughout the cosmic
history. However, in this paper we concentrate purely on the
potential variability of the most genuine fundamental
gravitational parameters of Einstein's equations such as $\CC$
and $G$. The possibility that $G$ could be associated to a
variable scalar field stems from the old Jordan and Brans-Dicke
proposals\,\cite{JBD} and has generated an abundant literature
since then. Similarly the idea that the $\CC$ term could perhaps
be variable and related to a dynamical field is also relatively
old\,\cite{Dolgov,PSW,Wetterich}. In short, whether related to
dynamical fields or to the general phenomenon of RG running, the
so-called ``fundamental constants of Nature'' are suspicious of
being non-constant from the point of view of quantum theory.
Adopting this general Ansatz, we wish to investigate whether
general (model-independent) properties can be extracted if we
take Einstein's equations (\ref{EE}) in the FRW metric, with
variable parameters (\ref{varibleCCG}), as the starting point for
the study of cosmology.}

Functions (\ref{varibleCCG}) will usually be monotonous; e.g. in
the RG model of Ref.\,\cite{RGTypeIa} $\rL$ inherits its
time/redshift dependence through $\rL=c_1+c_2\,H^2$ and it
satisfies $d\rL/dz\gtrless 0$ if $c_2\gtrless 0$ respectively.
{Similarly, $G=G_0/(1+\nu\ln H^2/H_0^2)$ in the RG model of
\cite{SSS}, and therefore $dG/dz\gtrless 0$ if $\nu\lessgtr 0$
respectively. Furthermore, in most phenomenological models in the
literature (where the underlying dynamics is usually not
specified) this monotonic character also
applies\,\cite{CCvariable1}-\cite{PSW}}. Since the functions
(\ref{varibleCCG}) are presumably known  from the fundamental
model, and the ordinary EOS for matter-radiation is also given,
one may solve Eq.\,(\ref{BianchiGeneral}) and Friedmann's equation
to determine $\rho=\rho(z),\,\rL=\rL(z)$ and $G=G(z)$ as explicit
functions of the redshift. These may then be substituted back in
(\ref{CCpicture}) to get the expansion rate of the variable $\CC$
model as a known function of $z$:
\begin{equation}\label{HLambda}
H_{\CC}^2(z)=H^2_0\,\left[\OM^0\,\fM(z;r)(1+z)^{\alpha}
+\OL^0\,\fL(z;r)\right]\,,
\end{equation}
where  $\alpha=3(1+\omega_m)$ ($\omega_m=1/3$ or $\omega_m=0$ for
the radiation or matter dominated epochs respectively). Functions
$\fM$ and $\fL$ are completely determined at this stage, and may
depend on some free parameters $r=r_1,r_2,..$ --
see\,\cite{RGTypeIa,SSS} for non-trivial examples with a single
parameter. Only for the standard $\CC CDM$ model we have
$\fM=\fL=1$. In general it is not so due to the mixed conservation
law (\ref{BianchiGeneral}). Whatever it be their form, these
functions must satisfy $\fM(0;r)=\fL(0;r)=1$ in order that the
cosmic sum rule $\OM^0+\OL^0=1$ is fulfilled.

On the other hand, within the context of the effective DE picture
the matter-radiation density satisfies by definition the standard
(``unmixed'') conservation law
\begin{equation}\label{scl}
\dot{\rs}+\alpha\,H_D\,\rs=0\,.
\end{equation}
Here we have denoted the matter-radiation density by $\rs$ to
emphasize that in the DE picture it satisfies the standard
self-conservation law, i.e. it corresponds to case i) mentioned
above. Moreover $H_D$ stands for
\begin{equation}\label{DEpicture}
H_D^2=\frac{8\pi G_0}{3}(\rs+\rD)
\end{equation}
with constant $G_0$. The subscript D here is to distinguish it
from the Hubble function in the CC picture, Eq.(\ref{CCpicture}).
The DE density $\rD$ is also conserved, independently of matter:
\begin{equation}\label{conservDE}
\dot{\rD}+3\,H_D(1+\we)\,\rD=0\,.
\end{equation}
The solutions of the two conservation equations (\ref{scl}) and
(\ref{conservDE}) read $\rs(z)=\rs(0)\left(1+z\right)^{\alpha}$
and
\begin{eqnarray}\label{zeta}
&&\rD(z)=\rD(0)\,\zeta(z)\,,\\
&&\zeta(z)\equiv\,\exp\left\{3\,\int_0^z\,dz'
\frac{1+\we(z')}{1+z'}\right\}\nonumber\,.
\end{eqnarray}
From (\ref{zeta}) it is clear that if $\we(z)\gtrsim-1$, then
$\rD$ increases with $z$ (hence decreases with the expansion).
This is characteristic of standard
quintessence\,\cite{QEpopular}. In contrast, if at some point
$z=z_p$ we hit $\we(z_p)<-1$, the DE density will decrease with
$z$ and hence will increase with the expansion. This anomaly
would signal the breakthrough of a phantom regime\,\cite{phantom}
in some interval of $z$ around $z_p$. The general solution for
$\rs(z)$ and $\rD(z)$ given above can be substituted back to
Friedmann's equation in the DE picture and one gets the
corresponding expansion rate:
\begin{eqnarray}\label{HzSS}
H_D^2(z) &=&
H^2_0\,\left[\tOM^0\,\left(1+z\right)^{\alpha}+\tOD^{0}\,\zeta(z)\right]\,.
\end{eqnarray}
Here $\tOM^0+\tOD^{0}=1$ because $\zeta(0)=1$. In general
$\Delta\Omega_M\equiv \OM^0-\tOM^0$ is nonzero. In fact, suppose
that one fits the high-z supernovae data using a variable $\rL$
model -- cf. the second Ref.\cite{{RGTypeIa}}. The fit crucially
depends on the luminosity distance function, which is determined
by the explicit structure of (\ref{HLambda}), so that the fitting
parameters $\OM^0,\OL^0$ can be different from those obtained by
substituting the alternate function (\ref{HzSS}) in the
luminosity distance.

\vspace{0.7cm}

\noindent {\bf 3. Effective EOS for a general model with variable
{$\CC$ and $G$}}\quad \vskip 0.4cm

Let us now compute $\we$. With the help of (\ref{zeta}) we have
$\we(z)=-1+(1/3)\,(({1+z)}/{\zeta})\,{d\zeta}/{dz}$.
Differentiating Eq.\,(\ref{HzSS}) on both sides we can eliminate
$d\zeta/dz$. Next we apply the matching condition mentioned
above, meaning that we fully identify $H^2_D$ with $H^2_{\CC}$ in
the effective DE picture. The final result is
\begin{equation}\label{wpzeta}
\we(z)=\frac{-1+((1+z)/3)\,
(1/\,H_{\CC}^2(z))\,d\,H_{\CC}^2(z)/dz+(1-\alpha/3)\,\tOM^0\,
\left(H^2_0/H^2_{\CC}(z)\right)\,(1+z)^{\alpha}}
{1-\tOM^0\,\left(H^2_0/H^2_{\CC}(z)\right)\,(1+z)^{\alpha}}\,.
\end{equation}
This is the effective barotropic index for a variable $\rL(z)$
and/or $G(z)$ model. The procedure is formally similar to
Ref\,\cite{Saini} but it is conceptually different. Here we do
not try to reconstruct the scalar field dynamics as a fundamental
model of the DE, not even a model-independent polynomial
approximation to the DE\,\cite{Alam}; rather, we start from a
general (purportedly fundamental) $\rL$ model that can decay into
matter and radiation and then we simulate it with a (generic)
dynamical DE model in which the matter-radiation is strictly
conserved, as usually assumed. The rationale for this is that
fundamental physics (e.g. the RG in QFT) can offer us useful
information on the possible functional forms (\ref{varibleCCG}),
as shown in\,\cite{JHEPCC1}-\cite{Reuter}. As a consequence,
$H_{\CC}(z)$ can actually be computed after solving the coupled
system of equations formed by Friedmann's equation, the
non-trivial conservation law (\ref{BianchiGeneral}) and the
relations (\ref{varibleCCG}) provided by the fundamental model.
Then one can check whether the effective EOS for the variable
$\rL$ model does emulate quintessence or phantom energy depending
on whether the calculation of the \textit{r.h.s.} of
(\ref{wpzeta}) yields $-1<\we<-1/3$ or $\we<-1$ respectively.

It is remarkable that the tracking of the $\rL(t)$ model by the DE
picture (in particular a possible phantom behavior near our time)
is actually expected for a wider class of variable $\rL$ models.
It is not easy to see this from Eq.(\ref{wpzeta}). Instead, let
us return to Eq.(\ref{BianchiGeneral}). From the aforementioned
matching condition we have $G(\rho+\rL)=G_0(\rs+\rD)$. Using also
$H\,dt=-dz/(1+z)$ we can transform the general Bianchi identity
(\ref{BianchiGeneral}) into the following differential form
\begin{equation}\label{droro}
{(1+z)}\,\,d(\rs+\rD)=\alpha\,\left(\rs+\rD-\xL\right)\,{dz}\,,
\end{equation}
where we have defined
\begin{equation}\label{xiL}
\xL(z)=\frac{G(z)}{G_0}\,\rL(z)\,.
\end{equation}
This new variable coincides with $\rL$ only if $G$ has the
constant value $G_0$, but in general both $\rL$ and $G$ will
develop a time/redshift evolution from the fundamental theory (for
instance from given RG equations), so that $\xL$ will be a known
function $\xL=\xL(\rho,H,...)$. Using (\ref{scl}), rewritten in
terms of the variable $z$, we can eliminate $\rs$ from
(\ref{droro}) and we are left with a simple differential equation
for $\rD$:
\begin{equation}\label{drdz}
\frac{d\rD(z)}{dz}=\alpha\,\frac{\rD(z)-\xL(z)}{1+z}\equiv\beta(\rD(z))\,.
\end{equation}
This equation is particularly convenient to analyze the behavior
of the effective DE picture. Consider the plane
$(\rD,\beta(\rD))$. For constant $\xL$ (as in the $\CC$CDM model,
where $G=G_0$ and $\xL=\rL^0$), Eq.\,(\ref{drdz}) resembles a
renormalization group equation for the ``coupling'' $\rD$, with
``$\beta$-function'' $\beta(\rD)$ and a fixed point (FP) at
$\rD^{*}=\xL$, where $\beta(\rD^{*})=0$. Since
$d\beta(\rD)/d\rD=\alpha/(1+z)>0$ (for all $z>-1$) it is clear
that this FP is an infrared (IR) stable fixed point, hence
$\rD(z)\rightarrow\rD^{*}$ for decreasing $z$. For variable
$\xL=\xL(z)$ the conclusion is the same: the solution $\rD=\rD(z)$
of (\ref{drdz}) is attracted to the given function $\xL=\xL(z)$ in
the far IR. The quintessence regime occurs for $\beta(\rD)>0$,
i.e. for $\rD>\xL$, whereas the phantom regime is characterized by
$\beta(\rD)<0$. Whatever it be the sign of $\beta(\rD)$ at a
particular instant of time in the history of the Universe, the
function $\rD(z)$ will eventually be driven to $\xL(z)$ for
decreasing $z$ (namely by the expansion of the Universe). Let us
next study if the two functions $\rD(z)$ and $\xL(z)$ can be very
close at some point near our time, and what are the
phenomenological implications. The differential equation
(\ref{drdz}) can be integrated in closed form for any $\xL(z)$:
\begin{equation}\label{IF}
\rD(z)=\left(1+z\right)^{\alpha}\left[\rD(0)-\alpha
\int_0^z\frac{dz'\,\xL(z')}{(1+z')^{(\alpha+1)}}\right]\,.
\end{equation}
Expanding $\xL(z)$ around $z=-1$ one can check that
$\rD(z)\rightarrow\xL(z)$ at sufficiently late time (i.e.
$z\rightarrow-1$). Finally, from (\ref{zeta}) and with the help
of (\ref{drdz}) we find
\begin{equation}\label{we2}
\we(z)=-1+\frac{\alpha}{3}\,\left(1-\frac{\xL(z)}{\rD(z)}\right)\equiv-1+\epsilon(z)\,,
\end{equation}
where $\rD$ is given by (\ref{IF}). This formula is more useful
than Eq.(\ref{wpzeta}) in the present context. It provides an
efficient recipe to compute $\we$ directly from the sole knowledge
of $\xL$. Using these general formulae we can e.g. reproduce the
effective EOS obtained for the particular model recently studied
in Ref.\cite{SS1}, see Eq.(23-24) of the latter. At the present
time (that is, near $z=0$) we expect that $\epsilon(z)$ is a
relatively small quantity because observations show that the
effective barotropic index must be near $-1$ at
present\,\cite{Supernovae,WMAP03,LSS}. The sign of $\epsilon(z)$
at $z=0$, however, is not known with certainty. Model independent
analyses of the most recent cosmological data seem to
indicate\,\cite{Alam,Jassal} that $\we$ has undergone a certain
evolution from $\we\lesssim 0$ (at $z\simeq 1.7 $) reaching the
phantom regime $\we\lesssim -1$ at $z\gtrsim 0$. In \,\cite{SS1}
it was shown that a RG model for the $\rL$ evolution can exhibit
this type of behavior. We will now prove that this feature
actually holds for a large class of models with variable $\xL$.

\vspace{0.7cm}

\noindent {\bf 4. Effective quintessence/phantom behavior of
models with variable {$\CC$ and $G$} }\quad \vskip 0.4cm

Let us rewrite the solution of the differential equation
(\ref{drdz}) in the following alternative way:
\begin{equation}\label{IF2}
\rD(z)=\xL(z)-\left(1+z\right)^{\alpha}\,
\int_{z^{*}}^z\frac{dz'}{(1+z')^{\alpha}}\frac{d\xL(z')}{dz'}\,,
\end{equation}
where $z^{*}$ is a root of $\beta(\rD(z))=0$, i.e. a point where
$\rD(z^{*})=\xL(z^{*})$. Remarkably, one can show that a root
$z^{*}$ \textit{always} exists near our present time, meaning in
our recent past, in the immediate future or just at $z^{*}=0$. The
proof is based on establishing the following relation:
\begin{equation}\label{dzeta}
\frac{d\zeta(z)}{dz}=\frac{\alpha\,(1+z)^{\alpha-1}}{1-\tOM^0}
\left(\OM^0\,\fM(z;r)-\tOM^0\right)\,.
\end{equation}
It ensues after a straightforward calculation from: i) the
matching condition of the two pictures,
$H_{\Lambda}(z)=H_{D}(z)$, and ii) the constraint imposed on the
functions $\fM$ and $\fL$ in (\ref{HLambda}) by the Bianchi
identity (\ref{BianchiGeneral}). Since $\fM(0;r)=1$, as noted
previously, it is clear that if the cosmological parameters of the
two pictures coincide ($\Delta\OM=0$) the derivative (\ref{dzeta})
vanishes identically at $z^{*}=0$, hence $\we(0)=-1$. If, however,
$\Delta\OM\neq 0$ but is small (after all $\tOM^0$ and $\OM^0$ in
the two pictures should not be very different), then there exists
a point $z^{*}\gtrless 0$ not very far from $z=0$ where
$d\zeta/dz|_{z=z^{*}}=0$ (and so $\we(z^{*})=-1$). This completes
the proof. Therefore, $z^{*}\simeq 0$ exists and defines a
(local) divide between a quintessence phase and a phantom phase.
The slope of the function $\rD$ reads
\begin{equation}\label{dIF2}
\frac{d\rD(z)}{dz}=-\alpha\,\left(1+z\right)^{\alpha-1}
\int_{z^{*}}^z\frac{dz'}{(1+z')^{\alpha}}\frac{d\xL(z')}{dz'}\,.
\end{equation}
One could naively think that for \textit{increasing/decreasing}
$\xL$ with redshift, $\we(z)$ should always be
\textit{above/below} $-1$. Let us assume that $\xL$ is monotonous
(hence $d\xL/dz$ has a definite sign) in some interval containing
$z^{*}$ and some reference point $z=z_1$. Eq.\,(\ref{dIF2}) shows
that in this case $d\rD/dz$ is also monotonous, and if $z^{*}<z_1$
then $d\rD/dz|_{z=z_1}$ has opposite sign to $d\xL/dz|_{z=z_1}$.
Thus e.g. if $d\xL/dz>0$ (i.e. for $\xL$ decreasing with the
expansion) then $d\rD/dz|_{z=z_1}<0$, meaning that  the dynamical
DE picture will look as phantom energy at $z_1$ (contrary to
na\"ive expectations); if $z^{*}>z_1$ then $d\rD/dz|_{z=z_1}>0$
and the DE will look as  quintessence at $z_1$. Similarly, if
$\xL(z)$ is monotonically increasing with the expansion
($d\xL/dz<0$), and $z^{*}<z_1$, the observer at $z_1$ will see
quintessence (again counterintuitive), but when $z_1<z^{*}$ he/she
will see phantom DE. The last situation is particularly worth
noticing because if $z_1=0$ this case could just correspond to the
present observational data (if the tilt into the phantom regime is
finally confirmed\,\cite{Alam,Jassal}). The exact position of
$z^{*}$ depends on the peculiarities of the variable $\xL=\xL(z)$
model and also on the values of the parameter difference
$\Delta\OM$ in the two pictures (\ref{HLambda}) and (\ref{HzSS}).
This is corroborated in concrete scenarios with monotonically
decreasing $\rL(z)$ \,\cite{SS1}, where for parameter differences
of a few percent one finds that the effective EOS develops a
phantom phase at $z^{*}\gtrsim 0$ and persists phantom-like
asymptotically until the remote future, where $\we(z)\rightarrow
-1$.

Finally, we note from (\ref{BianchiGeneral}) that if $G=G(t)$ and
we assume that  matter is conserved (no transfer of energy with a
variable $\rL$), then ${d\xL}/{dt}=-(\rho/G_0)\,{dG}/{dt}$. In
this case if $G(t)$ is monotonous, $\xL(t)$ is too. From
(\ref{dIF2}) we get
\begin{equation}\label{varG}
\frac{d\rD}{dz}=\alpha (1+z)^{\alpha-1} \,\frac{\rho(0)}{G_0}\
[G(z)-G(z^{*})]\,.
\end{equation}
This shows that if $G$ is asymptotically free -- that is to say,
if $G$ decreases with redshift -- we should observe quintessence
behavior for $0\leqslant z\leqslant z^{*}$; whereas if $G$ is
``IR free'' (i.e. it increases with $z$ -- hence decreases with
the expansion), then we should have phantom behavior in that
interval. In general, if $\rL$ is variable (and $G$ is fixed or
variable) the monotonous property of $\xL$ is satisfied by most
models in the literature\,\cite{CCvariable2}, including the RG
ones\,\cite{JHEPCC1,Babic,RGTypeIa,SSS,Reuter}. The monotonous
variation of  $\xL(z)$ around $z^{*}$ is important to insure a
long quintessence-like regime preceding this point -- as
suggested by the cosmological data.

\vspace{0.7cm}
 \noindent {\bf 5. An example: renormalization
group model with running $\rL$}\quad \vskip 0.4cm

{As a concrete example of cosmological model with variable
parameters, consider the case $G=const.$ and $\dot{\rL}\neq 0$. We
have catalogued this possibility in Sect.2 as case ii)}, for which
the general Bianchi identity (\ref{BianchiGeneral}) simplifies
into
\begin{equation}\label{Bronstein}
\dot{\rL}+\dot{\rho}+3\,H\,(\rho+p)=0\,.
\end{equation}
This equation, in combination to Friedmann's equation, has been
used to solve the cosmological RG model of \cite{RGTypeIa}. This
model treats the CC semiclassically as a running parameter,
therefore one evolving with time/redshift due to the quantum loop
effects of the high energy fields (the only ones that can
contribute significantly in this model, due to the
``soft-decoupling'' phenomenon associated to the CC in
Ref.\,\cite{RGTypeIa}). Interestingly enough this model can
simulate an apparent phantom behavior near our time\,\cite{SS1},
actually following very closely the polynomial data fits of
Ref.\,\cite{Alam}. In this particular model equations
(\ref{varibleCCG}) read
\begin{equation}\label{CCH}
\rL(z)=C_0+C_1\,H^2(z)\,,\ \ \ \ G={\rm const}\,,
\end{equation}
with
\begin{equation}\label{C0C1}
C_0=\rho_{\Lambda,0}-\frac{3\,\nu}{8\pi}M_P^2\,H_0^2\,, \ \ \
C_1=\frac{3\,\nu}{8\pi}\,M_P^2\,.
\end{equation}
This model has a single parameter, $\nu$, defined essentially as
the ratio (squared) of the masses of the high energy fields to the
Planck mass ($M_P$) \,\cite{RGTypeIa},
\begin{equation}\label{nu}
\nu=\frac{\sigma}{12\pi}\frac{M^2}{M_P^2}\,.
\end{equation}
Here $\sigma=\pm 1$ depending on whether bosons or fermions
dominate in their loop contributions to the running of $\rL$. The
typical value for $\nu$ is obtained when $M=M_P$, namely
\begin{equation}\label{nu0}
\nu_0\equiv\frac{1}{12\pi}\simeq 0.026\,.
\end{equation}
In general we expect $|\nu|\leqslant\nu_0$ because from the
effective field theory point of view we assume $M\leqslant M_P$.
This is also suggested from the bounds on $\nu$ obtained from
nucleosynthesis\,\cite{RGTypeIa} and also from the
CMB\,\cite{WangOpher}.

Using equations (\ref{Bronstein}) and (\ref{CCH}) in combination
with Friedmann's equation one can solve this cosmological model
explicitly and one can obtain the functions $H=H(z;\nu)$,
$\rho=\rho(z;\nu)$ and $\rL=\rL(z;\nu)$ in close analytic form.
We refer the reader to\,\cite{RGTypeIa} for all the details
concerning this running CC model. Once the model is solved we
have an explicit expression for the function $\xL(z)$,
Eq.(\ref{xiL}), which in this particular case is obviously given
by $\xL(z)=\rL(z;\nu)$. For the flat case one finds (in the
matter-dominated epoch, $\alpha=3$)
\begin{equation}\label{xL2}
\xL(z;\nu)=\rL^0+\rM^0\,\frac{\nu}{1-\nu}\,\left[\left(1+z\right)^{3(1-\nu)}-1\right]\,,
\end{equation}
where $\rL^0$ and $\rM^0$ are the values of these parameters at
$z=0$. Notice that for $\nu=0$ we recover the standard FRW case
with constant cosmological term. Substituting (\ref{xL2}) in the
general formula (\ref{IF}), and integrating, one obtains the
effective DE density associated to this model:
\begin{eqnarray}\label{effDE}
\rD(z;\nu)=\rL^0+(\rD^0-\rL^0)\,(1+z)^3
+\frac{\rM^0}{1-\nu}\,(1+z)^3\left[(1+z)^{-3\nu}-1\right]+
\frac{\nu\rM^0}{1-\nu}\left[(1+z)^{3}-1\right]\,.
\end{eqnarray}
As expected from the general discussion below Eq.(\ref{drdz}) we
confirm that $\rD(z;\nu)\rightarrow\xL(z;\nu)$ in the remote
future; in other words $\we(z\to -1)\rightarrow -1$, see
Eq.\,(\ref{we2}). From (\ref{xL2}) and (\ref{effDE}) we have the
asymptotic limit
\begin{equation}\label{asimpt}
\rD(z;\nu)\rightarrow\xL(z;\nu)\rightarrow\rL^0-\frac{\nu\,\rM^0}{1-\nu}\,\
\ \ ({\rm for}\ z\rightarrow -1)\,.
\end{equation}
\begin{figure}[t]
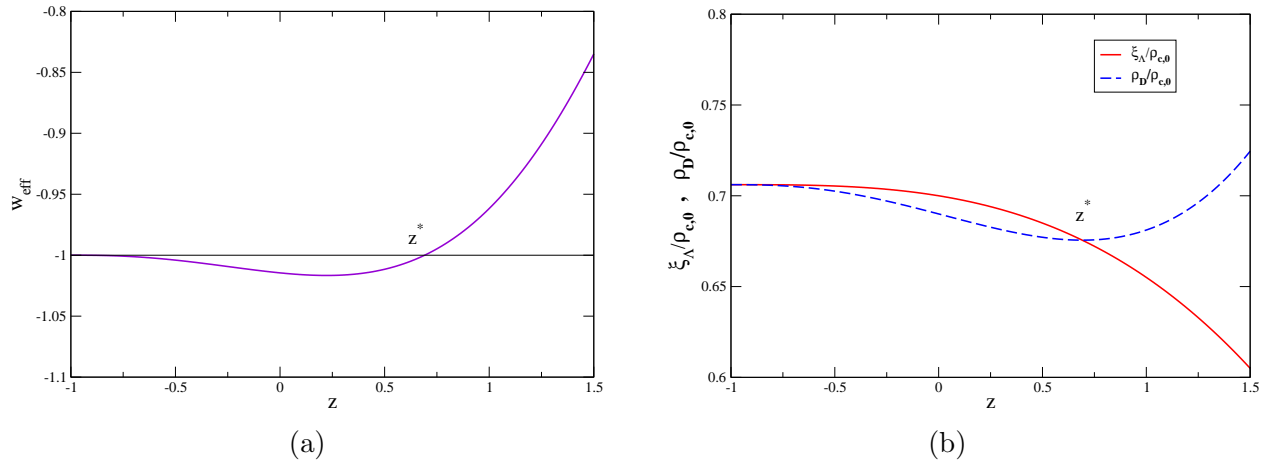

    \begin{tabular}{cc}
      \resizebox{0.48\textwidth}{!}{\includegraphics{fig1a.eps}} &
      \hspace{0.3cm}
      \resizebox{0.48\textwidth}{!}{\includegraphics{fig1b.eps}} \\
      (a) & (b)
    \end{tabular}
\caption{\textbf{(a)}\ Numerical analysis of the effective EOS
parameter $\we$, Eq.\,(\protect\ref{we2}), as a function of the
redshift for fixed $\nu=-0.8\,\nu_0<0$ and for
$\Delta\Omega_M=-0.01$. The Universe is assumed to be spatially
flat ($\OK^0=0$) with the standard parameter choice
$\OM^0=0.3\,,\OL^0=0.7$. In the figure, $z^{*}$ is the crossing
point of the ``barrier'' $\we=-1$; \textbf{(b)} The corresponding
evolution and crossing (at $z=z^{*}$) of the density functions
$\xL(z)$ and $\rD(z)$ in the two pictures, Eqs.
(\ref{xL2})-(\ref{effDE}), in units of the critical density
$\rho_{c,0}$ at present.}
  \label{plot}
\end{figure}
Finally, the effective EOS parameter for this model is obtained
by inserting (\ref{xL2}) and (\ref{effDE}) in Eq.(\ref{we2}). Upon
some rearrangement it finally yields
\begin{equation}\label{wpflat1}
\we(z;\nu)
=-1+(1-\nu)\,\frac{\OM^0\,(1+z)^{3(1-\nu)}-\tOM^0\,(1+z)^3}
{\OM^0\,[(1+z)^{3(1-\nu)}-1]-(1-\nu)\,[\tOM^0\,(1+z)^3-1]}\,,
\end{equation}
where we have defined $\OM^0=\rM^0/\rco$ and $\tOM^0=\rs^0/\rco$
corresponding to the matter densities in the two pictures. Recall
that in the CC picture $\rM$ is \textit{non}- conserved in this
model -- see Eq.(\ref{Bronstein}) -- whereas in the DE picture
$\rs$ is conserved -- cf\, Eq\,(\ref{scl}) --, so the two mass
density parameters $\rM^0$ and $\rs^0$ involved in the two fits
(in each picture) of the same data  need not to coincide. In our
general discussion below Eq.\,(\ref{HzSS}) we expressed this fact
by asserting that the parameter difference $\Delta\Omega_M\equiv
\OM^0-\tOM^0$ is in general expected to be nonzero\,\footnote{We
note that the result (\ref{wpflat1}), obtained from the general
method devised in this paper, does perfectly agree with the
particular calculation performed in \cite{SS1}.}.

In Fig.\,\ref{plot}a we show the numerical analysis of
Eq.\,(\ref{wpflat1}) for a typical choice of the parameters. In
general we do not expect $\Delta\Omega_M$ to be very large
because the two pictures are supposed to give a similar
representation of the same data. Therefore we have chosen
$\Delta\Omega_M$ of order one percent. Similarly, the parameter
$\nu$ should be close to (\ref{nu0}). For $\Delta\Omega_M<0$ and
$\nu<0$ as given in Fig.\ref{plot}a we find a transition point
from quintessence-like behavior into phantom-like behavior very
near our recent past, namely around $z^{*}\simeq 0.7$. -- marked
explicitly in the figure -- i.e. around the time when the
Universe deceleration changed into acceleration. In
Fig.\,\ref{plot}b we plot the density functions (\ref{xL2}) and
(\ref{effDE}) normalized to the critical density at present,
$\rco$. This plot also exhibits the crossing point $z=z^{*}$ of
the two curves, where $\xL(z^*)=\rD(z^*)$. As predicted by the
general differential equation (\ref{drdz}), in this figure the
quintessence-like behavior is seen to be characterized by
$\rD(z)>\xL(z)$ whereas the phantom-like behavior by
$\rD(z)<\xL(z)$, with $z=z^*$ acting as a transition point
between the two. We can also check in Fig.\,\ref{plot}b the
accomplishment of the asymptotic condition given by
Eq.\,(\ref{asimpt}).

The interesting features displayed by this running cosmological
model are only a particular illustration of the general
expectations for cosmological models with variable cosmological
parameters. The alternate representation of the effective DE
density $\rD(z)$ and its derivative $d\rD/dz$ -- see equations
(\ref{IF2}) and (\ref{dIF2}) -- allowed us to describe these
features in a completely general way without committing to any
particular model. If we change the model (namely the kind of
evolution of the redshift functions (\ref{varibleCCG})) similar
features will arise depending on the parameters of the model and
on the value of $\Delta\Omega_M$, but in all cases a transition
point $z^*$ near our present time (lying either in our recent
past or in our immediate future) will be found. The existence of
the crossing point $z^{*}$ ($\we(z^{*})=-1$) where the effective
EOS may change from quintessence-like to phantom-like behavior
near our present time is, as we have proven in this paper, a
general result for all the models with variable cosmological
parameters.

\vspace{0.5cm}

 \noindent {\bf 6. Conclusions}\quad

 \vspace{0.2cm}

To summarize, we have shown that a model with variable $\rL$
and/or $G$ generally leads to an effective EOS with a non-trivial
barotropic index, with the property $\we(z)\rightarrow -1$ in the
far IR. Such model can mimic a dynamical DE model and effectively
appear as quintessence and even as phantom energy. The eventual
determination of an empirical EOS for the DE in the next
generation of precision cosmology experiments should keep in mind
this possibility. The nature of the dynamics behind $\rL$ is not
known in principle, but various works in the literature suggest
that a fundamental $\rL$ can display a RG running which can be
translated into redshift evolution. Here we have generalized these
ideas and have shown that in \textit{any} model with variable
$\xL(z)=({G(z)}/{G_0})\,\rL(z)$ there is an effective DE density,
$\rD(z)$, that tracks the evolution of $\xL(z)$ and converges to
it in the far IR. Moreover, we have proven that there
\textit{always} exists a point $z^{*}$ near $z=0$ where
$\rD(z^{*})=\xL(z^{*})$, hence $\we(z^{*})=-1$. If this point lies
in our recent past ($z^{*}\gtrsim 0$) and $\xL$ is a decreasing
function of $z$ around $z^{*}$, then there must necessarily be a
recent transition into an (effective) phantom regime
$\we(z)\lesssim-1$. This would explain in a natural way another
``cosmic coincidence'': why the crossing of $\we=-1$ just occurs
near our present time? Our results are \textit{model-independent},
they only assume the formal structure of Einstein's equations of
General Relativity and FRW cosmology in the presence of a variable
cosmological term, $\rL=\rL(z)$, and possibly (though not
necessarily) a variable gravitational coupling $G=G(z)$. We
conclude that there is a large class of variable $(\rL,G)$ models
that could account for the observed evolution of the DE, without
need of invoking any combination of fundamental quintessence and
phantom fields. {Last, but not least, we wish to point out that
PLANCK and SNAP\,\cite{SNAP} data can constrain the underlying
model with varying parameters (e.g. the RG model of Sect. 5) --
see also\,\cite{RGTypeIa}. Since, in principle, the underlying
model may also have implications at astrophysical
scales\,\cite{SSS}, one could even check if the constraints from
SNAP and PLANCK can yield the expected behavior at these scales.
Most important, if the SNAP and PLANCK data should confirm the
crossing of the CC boundary, $\we(z^{*})=-1$, it would strengthen
this approach substantially since the  CC boundary crossing is
generic in the entire class of cosmological models with variable
cosmological parameters studied in this paper. Notice, however,
that if observations would prove that $\we>-1$ this would not
invalidate the model because the crossing point  could be in our
immediate future ($z^*\lesssim 0$). What we have proven, indeed,
is that in this kind of models there always exists a crossing
point $z^*$ around our present ($z=0$), whether in our recent
past or in our near future.}

 \vspace{0.7cm}

 \noindent {\bf Acknowledgments.}\quad

 \vspace{0.3cm}

We thank A. Dolgov and A.A. Starobinsky for useful discussions.
This work has been supported in part by MEC and FEDER under
project 2004-04582-C02-01; JS is also supported by Dep. de
Recerca, Generalitat de Catalunya under contract CIRIT GC
2001SGR-00065. The work of HS is financed by the Secretaria de
Estado de Universidades e Investigaci\'on of the Ministerio de
Educaci\'on y Ciencia of Spain. HS thanks the Dep. ECM of the
Univ. of Barcelona for the hospitality.

\newcommand{\JHEP}[3]{{\sl J. of High Energy Physics } {JHEP} {#1} (#2)  {#3}}
\newcommand{\NPB}[3]{{\sl Nucl. Phys. } {\bf B#1} (#2)  {#3}}
\newcommand{\NPPS}[3]{{\sl Nucl. Phys. Proc. Supp. } {\bf #1} (#2)  {#3}}
\newcommand{\PRD}[3]{{\sl Phys. Rev. } {\bf D#1} (#2)   {#3}}
\newcommand{\PLB}[3]{{\sl Phys. Lett. } {\bf #1B} (#2)  {#3}}
\newcommand{\EPJ}[3]{{\sl Eur. Phys. J } {\bf C#1} (#2)  {#3}}
\newcommand{\PR}[3]{{\sl Phys. Rep } {\bf #1} (#2)  {#3}}
\newcommand{\RMP}[3]{{\sl Rev. Mod. Phys. } {\bf #1} (#2)  {#3}}
\newcommand{\IJMP}[3]{{\sl Int. J. of Mod. Phys. } {\bf A#1} (#2)  {#3}}
\newcommand{\PRL}[3]{{\sl Phys. Rev. Lett. } {\bf #1} (#2) {#3}}
\newcommand{\ZFP}[3]{{\sl Zeitsch. f. Physik } {\bf C#1} (#2)  {#3}}
\newcommand{\IJMPA}[3]{{\sl Int. J. Mod. Phys. } {\bf A#1} (#2) {#3}}
\newcommand{\MPLA}[3]{{\sl Mod. Phys. Lett. } {\bf A#1} (19#2) {#3}}
\newcommand{\CQG}[3]{{\sl Class. Quant. Grav. } {\bf #1} (#2) {#3}}
\newcommand{\JCAP}[3]{{\sl JCAP} {\bf#1} (#2)  {#3}}
\newcommand{\APJ}[3]{{\sl Astrophys. J. } {\bf #1} (#2)  {#3}}
\newcommand{\AMJ}[3]{{\sl Astronom. J. } {\bf #1} (#2)  {#3}}
\newcommand{\APP}[3]{{\sl Astropart. Phys. } {\bf #1} (#2)  {#3}}
\newcommand{\AAP}[3]{{\sl Astron. Astrophys. } {\bf #1} (#2)  {#3}}
\newcommand{\MNRAS}[3]{{\sl Mon. Not.Roy. Astron. Soc.} {\bf #1} (#2)  {#3}}



\end{document}